\def\>{\vskip .12in}

\documentstyle[prl,aps,,epsfig]{revtex}

\begin{document}
\preprint{SLAC-PUB-8387}
\preprint{March 2000}
\preprint{T/E}
\title{
     $\sigma_{A}/\sigma_{d}$ at low $x$ and low $Q^2$\thanks{Work
supported by the National Science Foundation and the Department of Energy.}}
\author
{ Stephen Rock and Peter Bosted}
\address{
\baselineskip 14 pt
University of Massachusetts;  Amherst, MA 01003-4525}
\maketitle 
\begin{abstract}
 We have extracted ratios of cross sections for scattering electrons from
high mass targets compared to low mass targets in the region of 
$x \sim 0.02$ and 
$Q^2 \leq 1$ (GeV/c)$^2$ from SLAC experiments performed
over the past three decades.  Additional analysis was needed for
radiative corrections, target end caps and calibration runs.
We observe no significant difference in the nuclear ratio for low
$Q^2$ compared to results at $Q^2 \geq 1.$
\end{abstract}

An unexpected strong $A$,  $Q^2$, $x$ and $\epsilon$ dependence of
electron-Nucleus inclusive scattering  was  recently reported by the
HERMES collaboration \cite{HERMES} for $x <0.03$, $Q^2<1$ (GeV/c)$^2$
and $\epsilon<0.5$.  They used the
27 GeV HERA stored polarized electron beam and polarized and unpolarized
gas jet targets to measure exclusive and inclusive electron nucleus and 
nucleon scattering.
 The observed effect is largest near $x \sim 0.02$
and $Q^2 \sim 0.5$(GeV/c)$^2$ and $\epsilon \sim 0.5,$ where in the lab frame
 $x=Q^2/(2M\nu)$
is the fractional momentum of the struck quark, $Q^2$ is the momentum transfer
to the scattered electron, $\nu$ is the difference between the energy of the
incoming and scattered electron,  $\epsilon^{-1}  = 1+2(1+{\nu^2/Q^2})
\tan^2 (\theta/2)$, and $\theta$ is the angle of the scattered electron.
The EMC effect\cite{EMC} ( the  $A$ dependence of the
ratio of $\sigma_A/\sigma_d$)  had
been observed\cite{emcnmc} to be almost independent of $Q^2$  for
 $Q^2 \geq 1$ (GeV/c)$^2$ over a large range of $Q^2$ and $x$. The HERMES
data seem to indicate a large $Q^2$ dependence of this ratio for  
$x <0.03$, $Q^2<1$ (GeV/c)$^2$ and $\epsilon<0.5$ 
(dubbed the ''HERMES effect'').  Miller, Brodsky and
Karliner\cite{Miller} have explained this by coherent contributions from 
nuclear mesons.

\begin{figure}[p]
\vspace*{7.7in}
\hspace*{.45in}
\includegraphics{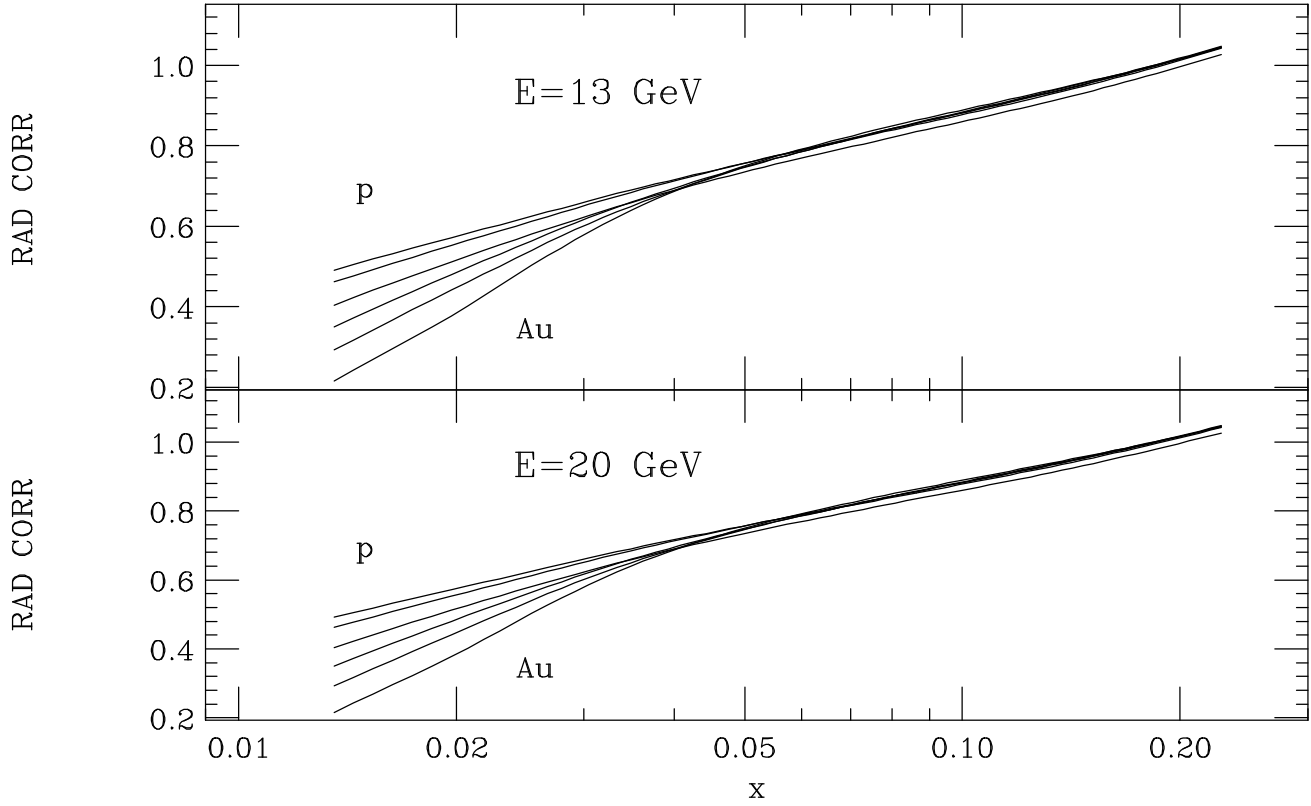}
\caption{Radiative Corrections for proton, deuteron, Be, Al, Cu and Au
for two different beam energies vs. $x$. At low $x$ the corrections are
large and different for the different materials due to the nuclear elastic 
tail. The labels 'p' and 'Au' indicate which curves correspond to which
nuclei (changing monotonically).}
\label{fg:rc}
\end{figure}

To further explore this kinematic region we have re-examined data from SLAC
experiments E61 \cite{Stein}, E140X \cite{E140X},
E154 \cite{E154} and E155X. Of these experiments, E61 comes closest to covering
the same kinematics as HERMES and uses a wide variety of nuclear targets.
The part of the data which is of
interest here had not been radiatively corrected
due to limitations in computing and theory in the early 1970s. 
Using more recent radiative corrections 
programs \cite{E140}, and the detailed table
of materials available in the paper, we have calculated the radiative
corrections over the entire measured region. 
 Most targets were 1\% radiators.  The radiative
corrections $\sigma_{Born}/\sigma_{measured}$ are shown in Fig. \ref{fg:rc}.
They are very large at low $x$ and depend strongly on the Atomic Number
because of the nuclear elastic tail. Using the Shell Model\cite{Shell}
 instead of the
Bessel Function Model\cite{Bessel} for the aluminum nuclear elastic cross 
section would
increase the Born cross section by 4\% at the lowest $x$ value at 20 GeV and
by 2\% at 13 GeV.  The errors on the Born cross section were calculated
using 
$\delta_{Born} =1 /(\sigma_{meas}/\sigma_{Born} -\sigma_{tail}/\sigma_{Born})
 \times \delta_{meas}$
where $\delta_{meas}$ is the experimental error on the measured cross section,
$\sigma_{tail}/\sigma_{Born}$ is the
sum of the nuclear elastic and quasi elastic tails.  This results
in a very large error for the gold target where these tails are a large 
fraction of the measured cross section at low $x$.

\begin{figure}
\vspace*{7.7in}
\hspace*{.45in}
\includegraphics{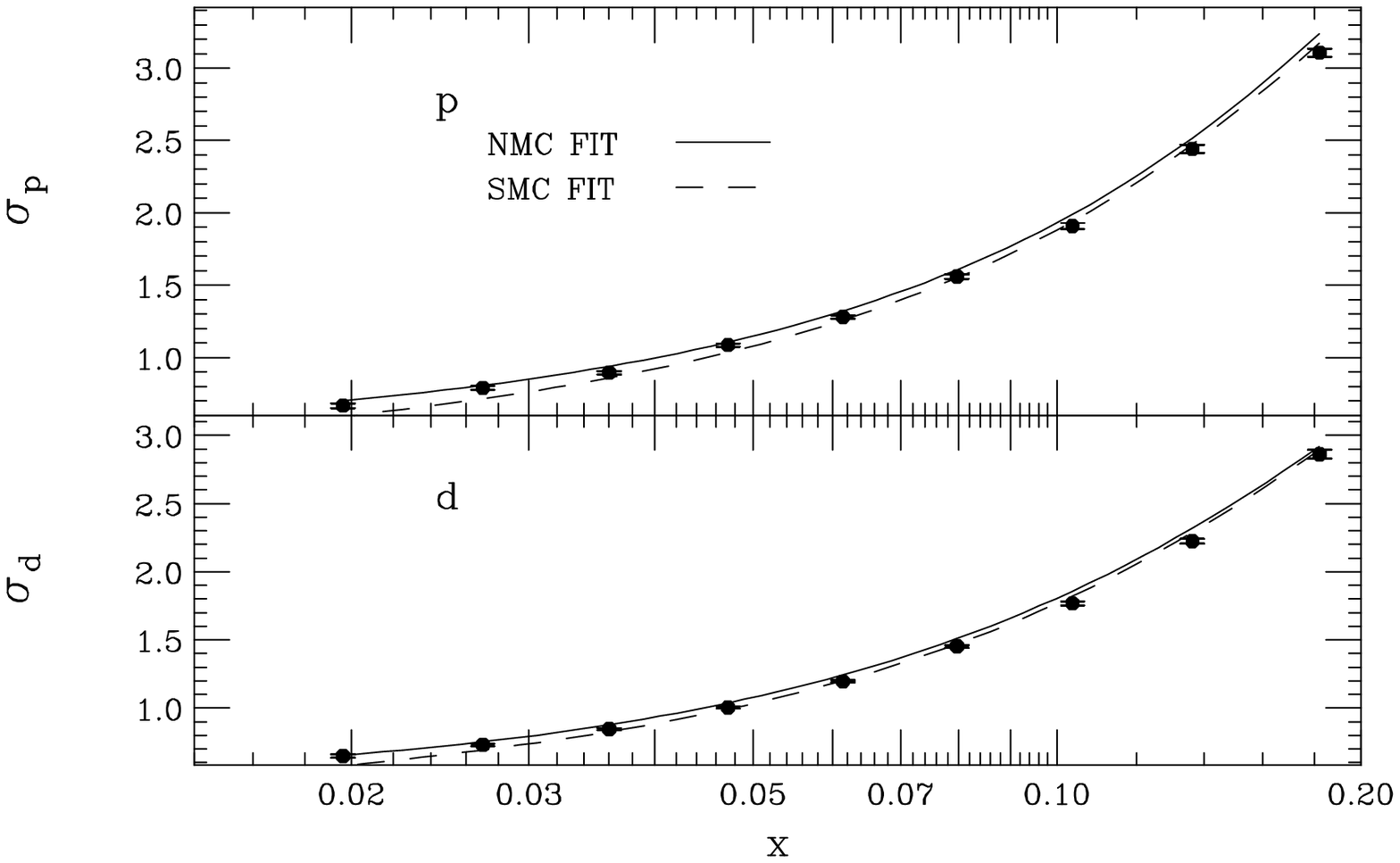}
\caption{Born cross sections for proton and deuteron (per nucleon) at 13 GeV
from Stein et al. compared to the NMC (solid) and SMC (dash) fit.  
These fits did not include
data in this region. The NMC fit is good, while the SMC fit is low at
low $x$.} 
\label{fg:cs13}
\end{figure}

\begin{figure}
\vspace*{7.7in}
\hspace*{.45in}
\includegraphics{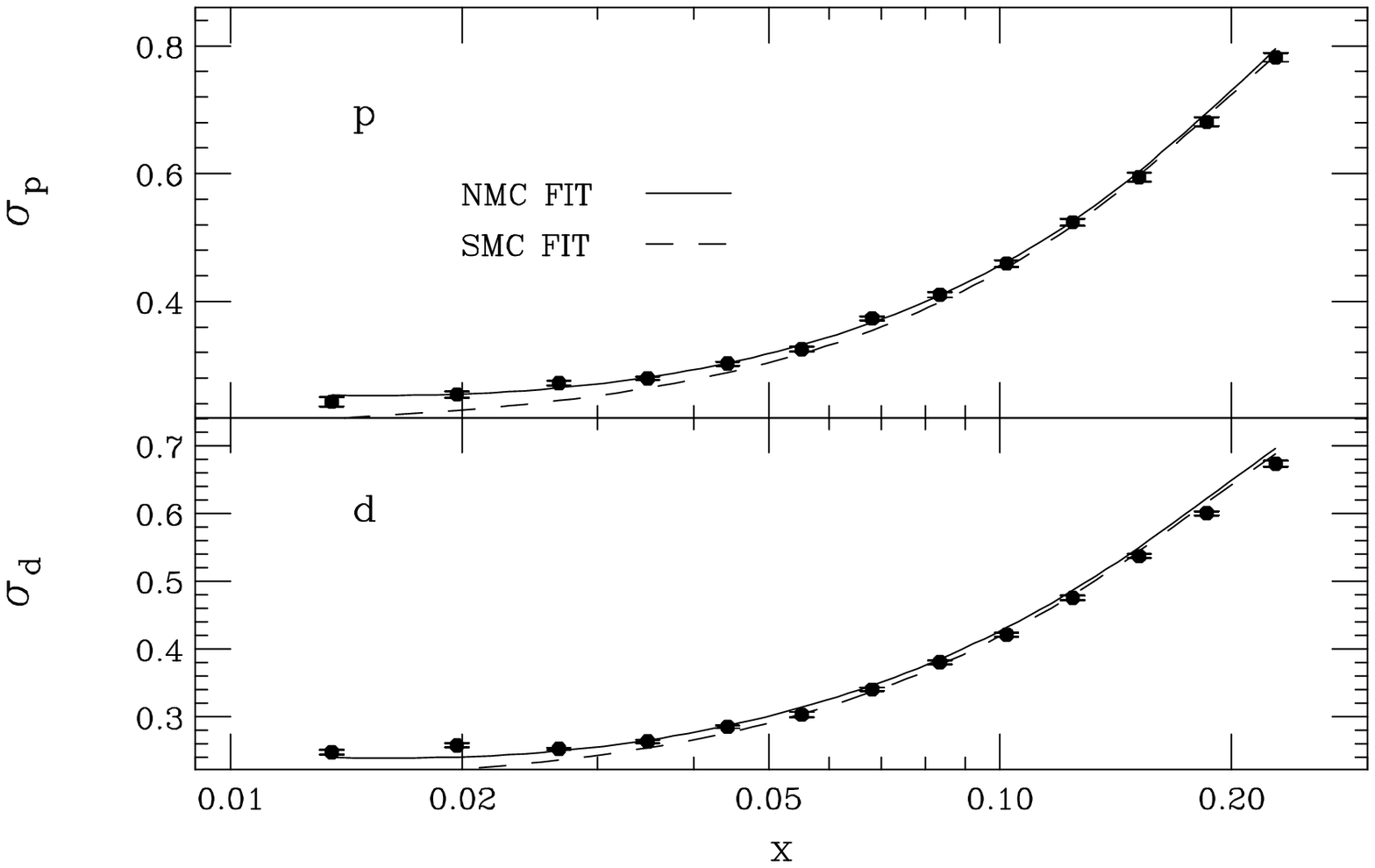}
\caption{Born cross sections for proton and deuteron (per nucleon) 
at E=20 GeV. from Stein et al. at 20 GeV compared to the 
NMC (solid) and SMC (dash) fit.  These fits did not include
data in this region. The NMC fit is good, while the SMC fit is low at
low $x$.} 
\label{fg:cs20}
\end{figure}

\begin{figure}
\vspace*{7.7in}
\hspace*{.45in}
\includegraphics{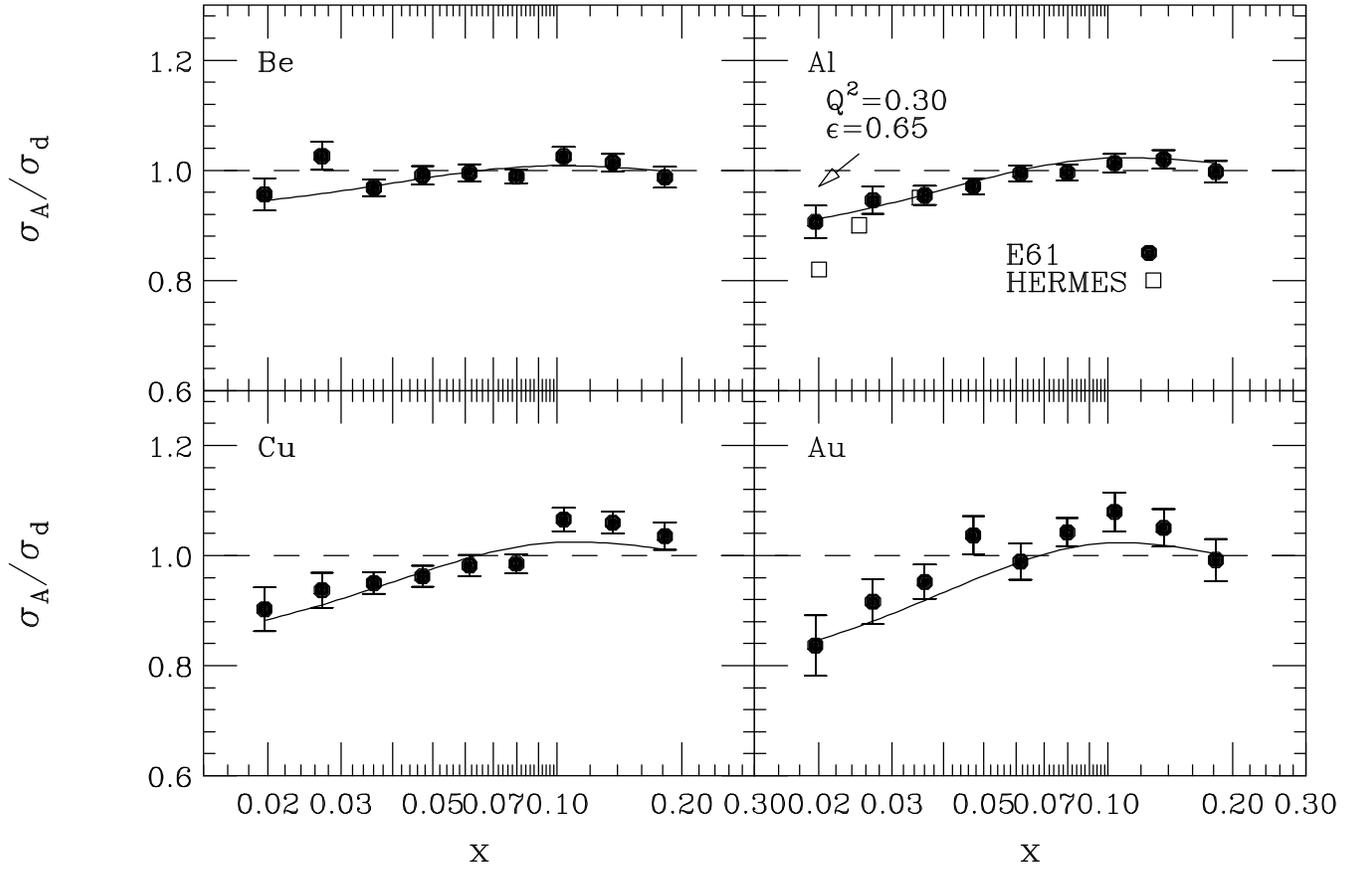}
\caption{$\sigma_A/\sigma_d$ as a function of $x$ for E=13 GeV.
The lowest $x$ points have ($Q^2,\epsilon$) = (0.3,0.65), (0.37,0.74).
The curve is the fit\protect\cite{emcfit} 
to EMC effect data at $Q^2>1$(GeV/c)$^2$.
Also shown with the aluminum results are the HERMES data\protect\cite{HERMES} 
for $\sigma_N/\sigma_d$ at the same $\epsilon$. The HERMES systematic error 
is  $\sim 0.03$. }
\label{fg:x13}
\end{figure}

\begin{figure}
\vspace*{7.7in}
\hspace*{.45in}
\includegraphics{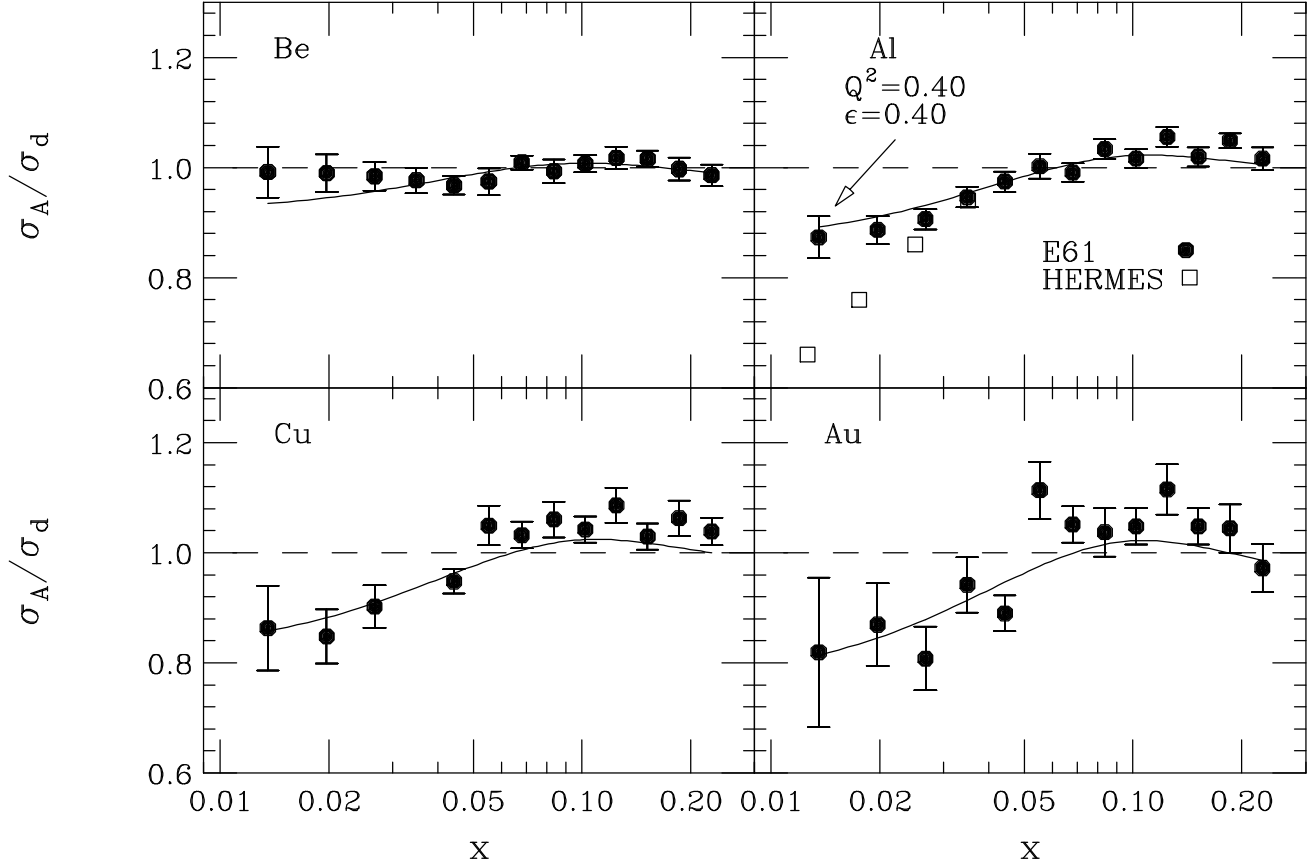}
\caption{$\sigma_A/\sigma_d$ as a function of $x$ for E=20 GeV.
The lowest $x$ points have ($Q^2,\epsilon$) =(0.40,0.40), (0.54,0.51), 
(0.66, 0.61.)
which overlap in $x$ and $\epsilon$ some of the HERMES data shown as open 
squares.  The HERMES systematic error is  $\sim 0.03$. 
The curve is the fit\protect\cite{emcfit} to EMC effect data at 
$Q^2>1$(GeV/c)$^2$.}
\label{fg:x20}
\end{figure}

\begin{figure}
\vspace*{7.7in}
\hspace*{.45in}
\includegraphics{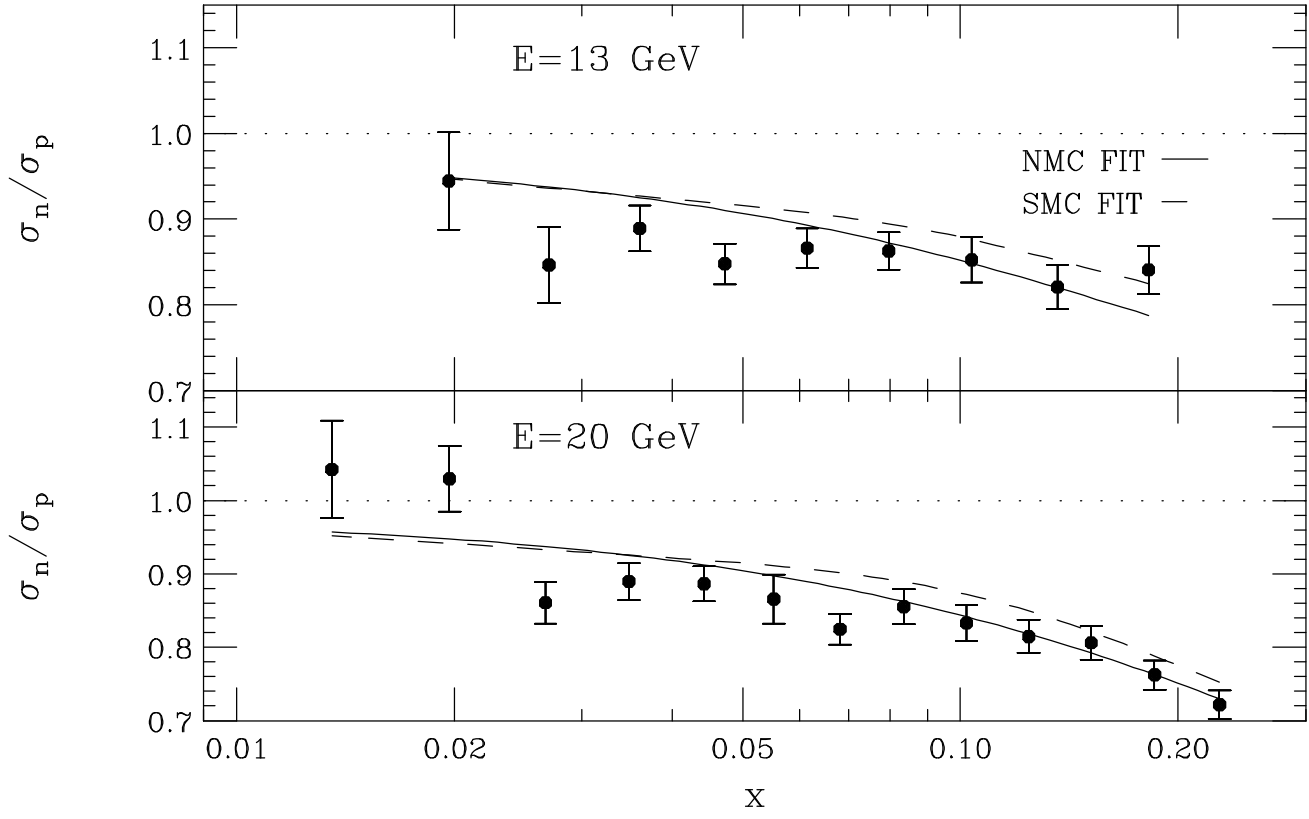}
\vskip -2.in 
\caption{$\sigma_n/\sigma_p$ as a function of $x$ for E=13 and 20 GeV.
Also shown are the NMC and SMC fits to $F_2^n/F_2^p$.}

\label{fg:np}
\end{figure}

 Figs \ref{fg:cs13} and  \ref{fg:cs20} show the E61 Born  cross sections 
for proton and deuteron (per nucleon) compared to the NMC\cite{NMC} and 
SMC\cite{SMC} fits using R1998 \cite{Fellbaum} for R.
The NMC fit is good while the SMC fit is low at low $x$. Note that these
fits did not include data in this kinematic region. 
 The ratio $\sigma_{A}/\sigma_{d}$ is shown in  Tables \ref{tb:x13} and 
\ref{tb:x20} and Figs. \ref{fg:x13} and \ref{fg:x20}. The solid line is
the fit in Gomez {\it et al.}\cite{emcfit} to data at higher $Q^2$ 
corrected for neutron excess. There is excellent agreement
between the data and the curves indicating little $Q^2$ dependence to the
$A$ dependence. The lower three $x$
points at $E=20$ GeV approximately overlap the kinematic region of the HERMES data 
in $x, Q^2$ and $\epsilon$.  HERMES data
are plotted in the panel for aluminum. They are at the same $\epsilon$ but
slightly higher $Q^2$.
There is a disagreement between the two data sets.
The ratio  $\sigma_{n}/\sigma_{p} = 2\times\sigma_{d}/\sigma_{p} -1$ is
shown in Fig. \ref{fg:np} for the two energies and the last 
column of Tables \ref{tb:x13} and \ref{tb:x20}. Also shown are the 
NMC fit \cite{npNMC} to $F_2^n/F_2^p$ and the 
results from the SMC fit. The NMC fit is in good agreement with the data
for most of the data set.  The curves may deviate slightly from the data
at the lowest values of $x$ at $E=20$ GeV.

SLAC experiment E140x \cite{E140X} measured $R=\sigma_L/\sigma_T$ for
protons and deuterons.  The target assemblies consisted of a
cylindrical aluminum cell with thin windows which contained the circulating 
$H_2$ or $D_2$ fluid, as well as an empty cell with thicker windows.
We have extracted the  ratio of cross sections from the aluminum to deuterium
as a function of $\epsilon$ to obtain the value of  $R_{Al} -R_d$ shown in 
Table \ref{tb:R}.
At  $x$=0.1 and $Q^2=0.5$(GeV/c)$^2$, the average ratio of cross sections is 
1.02 $\pm$ 0.007 and is independent of $\epsilon$ resulting in
 $R_{Al} -R_d$ = -0.033 $\pm$ 0.052, $R_{Al}/R_d$ = 0.9 $\pm$ 0.2. 
Comparison with  the  value of $R_{N}/R_d\sim 5$ reported by HERMES at 
$x\leq 0.03$ and the same $Q^2$ would require a very rapid increase in 
$R$  as a function of $x$.

\begin{figure}
\vspace*{7.7in}
\hspace*{.45in}
\includegraphics{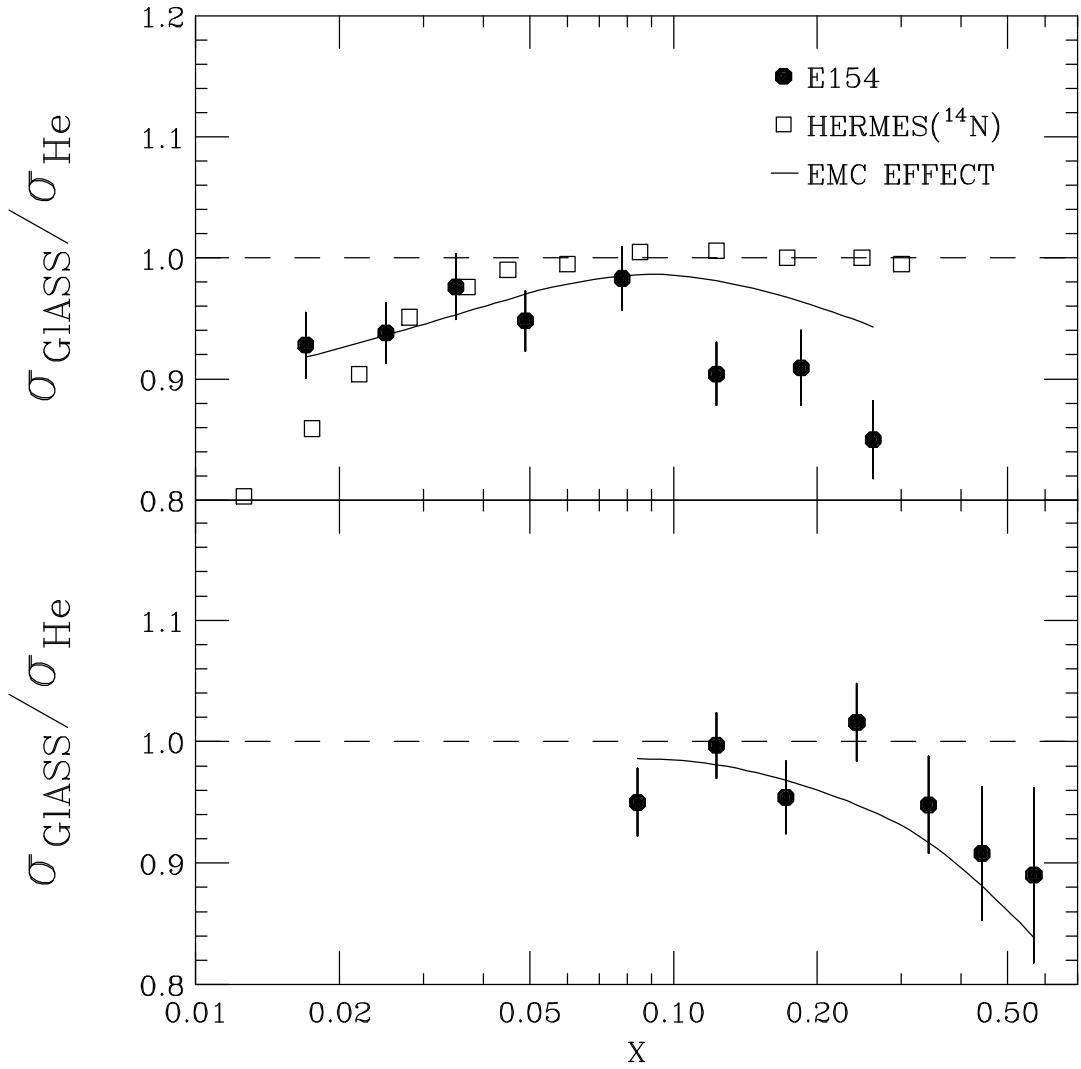}
\caption{
Measured ratio of cross sections   glass/$^3$He at
top) $2.75^o$ and bottom) $5.5^o$.  The E154
results (solid circle) are at $Q^2 > 1$(GeV/c)$^2$. The errors are the
total errors.
The three lowest HERMES\protect\cite{HERMES} points (open squares) are at
 $Q^2 < 1$ (GeV/c)$^2$. The HERMES systematic error is $\sim 0.03$
The solid curve is a fit to the EMC effect\protect\cite{emcfit}.
}
\label{fg:pol}
\end{figure}

The calibration data from E154 was used to determine the ratio
$\sigma_{glass}/\sigma_{^3He}$ in the kinematic range $x\geq  0.017$ and
$Q^2\geq 1.5$(GeV/c)$^2$ for $\epsilon \geq 0.4.$ This borders on the region where HERMES 
sees a large deviation from the EMC effect for the ratio of cross sections for
nitrogen to $^3He$.  The average $A$ for the glass used is $\sim 21$ compared
to $A=14$ for Nitrogen. The EMC effect depends logarithmically on $A$ and
if the HERMES effect does the same, there would be an enhancement in the 
effect of about 15\% of its value.
The experiment  \cite{E154} was designed to measure $g_1$ and $g_2$ for the 
neutron.
The polarized target was $^3He$ at 10 atmospheres pressure inside a 
glass cylinder.
The cylinder end caps and gas were of similar areal density. There was
also a calibration cell of almost identical design except that the 
gas pressure could be varied. Only ratios of counting rates were measured
since the acceptance of the spectrometer is not known to sufficient 
accuracy to determine useful absolute cross sections. All rates
were corrected for dead time and pion contamination. The dead time
was both rate and $x$ dependent, with a maximum of about 10\% at low
$x$ for a full target. An estimated uncertainty of 30\% in the dead time
correction corresponds to about 4\% systematic error on the counting rate
ratios. The counting rate from helium was determined by fitting the total 
counting rate
as a function of the pressure in the calibration cell. The counting rate from
glass was found by subtracting the pressure-normalized $^3$He counting rate
from the counting rate of the corresponding polarized cell. 
Four different calibration cells and 9 different polarized cells were used.
The ratio $\sigma_{glass}/\sigma_{^3He}$ was determined from
the ratio of counting rates and the measured 
dimensions of the polarized target, the helium density, the chemical 
composition of the glass, a fit to the ratio $\sigma_n/\sigma_p$ \cite{npNMC}, 
and radiative corrections.  The measurements on
the 9 targets scatter about the average with an width of less than 1 sigma. 
indicating that some of the errors, like radiative corrections, are
correlated between targets.
Fig. \ref{fg:pol} show  $\sigma_{glass}/\sigma_{^3He}$
averaged over all the measurements as a function of $x$ 
as measured by the spectrometers located
at $2.75^o$ and $5.5^o.$ The errors are statistical plus systematic. 
The results are consistent with a fit to the EMC effect\cite{emcfit}, 
except perhaps at the large $x$ at $2.75^o$.
Our lowest $x$ point is at $x=.017$, $Q^2=1.5$(GeV/c)$^2$ and $\epsilon=.37$.
The closest HERMES points (open squares) are at similar $x$ and $\epsilon$, 
but $Q^2 < 1$(GeV/c)$^2$ and have a systematic error of  $\sim 0.03$.
Within this systematic error E154 does not exclude the HERMES results within
the region of overlap.

\begin{figure}
\vspace*{7.7in}
\hspace*{.45in}
\includegraphics{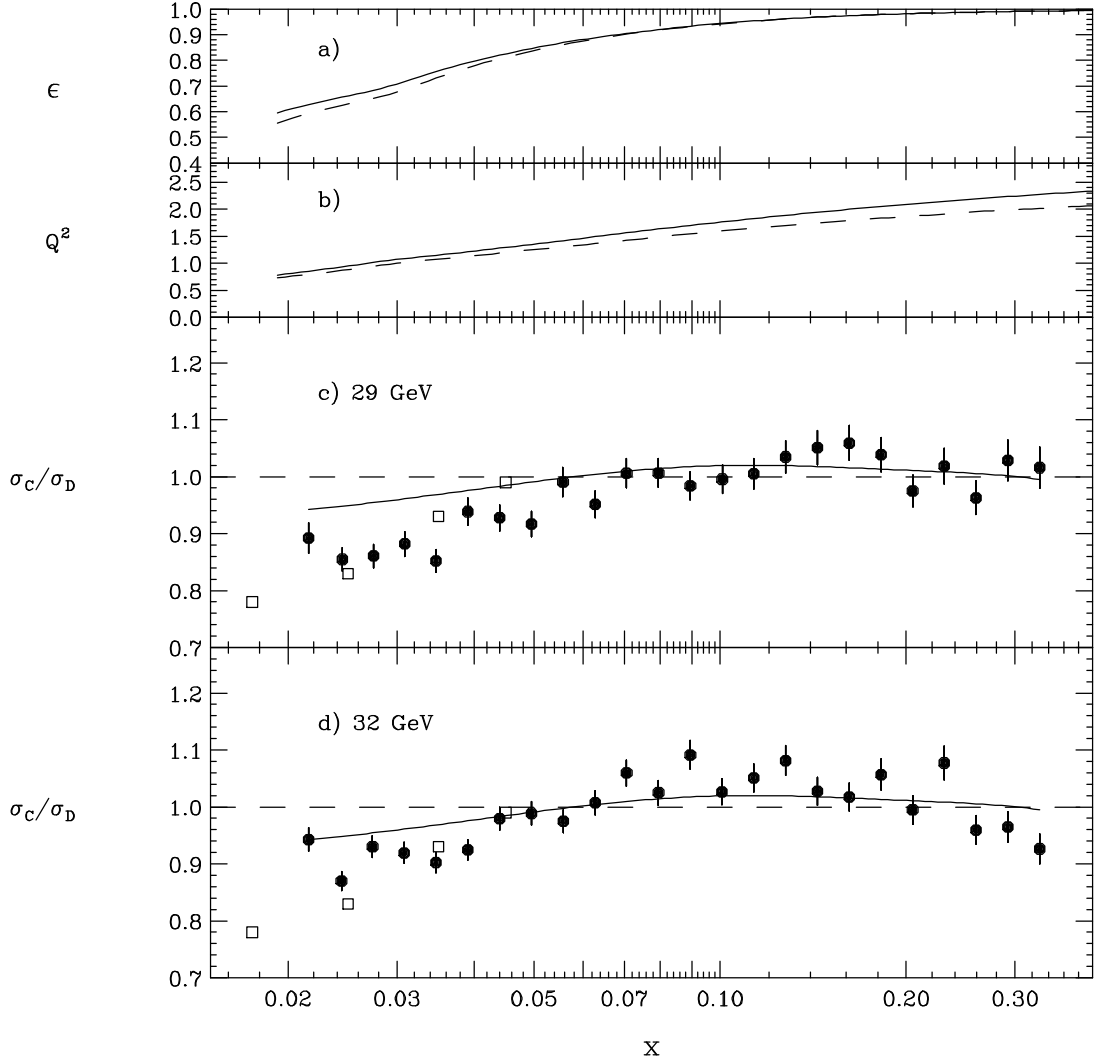}
\caption{Results from Experiment E155X on the Ratio $\sigma_C/\sigma_d$.
a,b) $\epsilon$ and $Q^2$ as a function of $x$ for the beam energies of
29 (dash) and 32 (solid) GeV. c,d) The ratio  $\sigma_C/\sigma_d$ from E155x
(solid circles), and from HERMES (open squares)\protect\cite{HERMES}. There
is a 10\% systematic error on the E155x results. The 
solid curve is a fit to the EMC effect at higher value of $Q^2$.}
\label{fg:e155x}
\end{figure}

 Experiment E155X used a NH$_3$ cryogenic polarized target to measure the
polarized structure function $g_2$.  The beam energies of 29 and 32 GeV
were approximately the same as HERMES. The spectrometers at 2.75$^o$ 
and 5.5$^o$ measured the scattered electrons. A carbon target was also used for
calibration.  From the ratio of the counting rates on carbon and
NH$_3$, we were able to determine the ratio $\sigma_C/\sigma_d$. The method
assumed that the cross sections for all the materials in the targets
(1.72 gms NH$_3$, 0.22 gms  $^4$He, 0.184 gms Al and 0.016 gms Cu or 
1.51 gms C, 0.39 gms $^4$He, and  0.18 gms Al) scaled in $A$ by the 
relative EMC 
effect \cite{emcfit}. The ratio was normalized to the predicted ratio
using the data from the 2.75$^o$ spectrometer for $x\geq 0.15$ and from the
5.5$^o$ spectrometer.  The $x$ dependent systematic errors come from radiative 
corrections, tracking efficiency, the mass of the Cu NMR coil in the NH$_3$ 
target and the uncertainty in $\sigma_p/\sigma_d$.  Because of the small fraction of Hydrogen in the polarized
target,  the error on $\sigma_C/\sigma_d$ is approximately 6 times the error
on $\sigma_{NH_3}/\sigma_C$, leading to  a systematic error of about 10\%.
Figs. \ref{fg:e155x}a and b show the $\epsilon$ and $Q^2$ values for the beam
energies of 29 GeV(dashed curve) and 32 GeV (solid curve) as a function of 
$x$. Unfortunately
the data do not extend down to $x\leq 0.02$ where the HERMES effect is
largest. Fig. \ref{fg:e155x}c and d show the 29 and 32 GeV cross section
ratios as the solid circles. The curve is the EMC effect \cite{emcfit} and
the open squares are the approximate values of the HERMES effect \cite{HERMES}
taken from their plots. The E155X data agree both with the EMC effect curve
and with the HERMES points within the 10\% systematic errors. 

In conclusion, SLAC data from several experiments are consistent with the
previously parameterized EMC effect and  constrain the
possibility of large effects as seen by HERMES. Experiment E61 used 
different target at
the same kinematics. E155X had similar kinematics, but large 
systematic errors. Experiments E140X and E154 used different high mass
targets at nearby kinematics.  Thus none of the experiments directly
contradicts the HERMES results on $\sigma_N/\sigma_d.$  However a smooth
interpolation between targets in E61 does not support the HERMES result.

\begin{table}
\caption{$R_{Al} -R_d$ and $R_{Al}/R_d$} 
\label{tb:R}
\begin{tabular}{rrrrr}
 $x$  & $Q^2$ & $R_{Al} -R_d$     & $R_{Al}/R_d$      & $\chi^2/df$ \\
\hline
 0.10  & 0.5 &  -0.033 $\pm$ .053   &  0.87 $\pm$ .20 & 0.5\\
 0.10  & 1.0 &   0.055 $\pm$ .118   &  1.16 $\pm$ .34 & 0.4\\
 0.35 & 3.0 &   0.074 $\pm$ .107   &  1.33 $\pm$ .46 & 0.8\\
 0.50  & 3.6 &  -0.011 $\pm$ .083   &  0.94 $\pm$ .43 & 4.5\\
\end{tabular}
\end{table}

\begin{table}
\caption{Ratio of Born cross sections for E=13 GeV} 
\label{tb:x13}
\begin{tabular}{rrrrrrrr}
 x  &  $Q^2$  & $\epsilon$ & Be/d  &  Al/d  & Cu/d  & Au/d & $\sigma_n/\sigma_p$ \\
 \hline
 0.182 &   0.69 &  0.98 &   0.99$\pm$0.02 &  1.00$\pm$0.02 & 1.03$\pm$0.02  &  0.99$\pm$0.04  &  0.84$\pm$0.03\\
 0.136 &   0.66 &  0.97 &   1.01$\pm$0.02 &  1.02$\pm$0.02 & 1.06$\pm$0.02  &  1.05$\pm$0.03  &  0.82$\pm$0.03\\
 0.104 &   0.62 &  0.96 &   1.03$\pm$0.02 &  1.01$\pm$0.02 & 1.07$\pm$0.02  &  1.08$\pm$0.04  &  0.85$\pm$0.03\\
 0.080 &   0.58 &  0.94 &   0.99$\pm$0.01 &  1.00$\pm$0.01 & 0.99$\pm$0.02  &  1.04$\pm$0.03  &  0.86$\pm$0.02\\
 0.061 &   0.53 &  0.91 &   1.00$\pm$0.02 &  0.99$\pm$0.01 & 0.98$\pm$0.02  &  0.99$\pm$0.03  &  0.87$\pm$0.02\\
 0.047 &   0.48 &  0.87 &   0.99$\pm$0.02 &  0.97$\pm$0.01 & 0.96$\pm$0.02  &  1.04$\pm$0.04  &  0.85$\pm$0.02\\
 0.036 &   0.43 &  0.81 &   0.97$\pm$0.02 &  0.95$\pm$0.02 & 0.95$\pm$0.02  &  0.95$\pm$0.03  &  0.89$\pm$0.03\\
 0.027 &   0.37 &  0.74 &   1.03$\pm$0.03 &  0.95$\pm$0.03 & 0.94$\pm$0.03  &  0.92$\pm$0.04  &  0.85$\pm$0.04\\
 0.020 &   0.30 &  0.65 &   0.96$\pm$0.03 &  0.91$\pm$0.03 & 0.90$\pm$0.04  &  0.84$\pm$0.05  &  0.94$\pm$0.06\\
\hline 

\hline
\end{tabular}
\end{table}

\begin{table}
\caption{Ratio of Born cross sections for E=20 GeV} 
\label{tb:x20}
\begin{tabular}{rrrrrrrr}
 x  &  $Q^2$  & $\epsilon$ & Be/d  &  Al/d  & Cu/d  & Au/d & $\sigma_n/\sigma_p$ \\
 \hline
 0.228 &   1.59 &  0.98 &   0.99$\pm$0.02 &  1.02$\pm$0.02 & 1.04$\pm$0.02  &  0.97$\pm$0.04  &  0.72$\pm$0.02\\
 0.186 &   1.52 &  0.97 &   1.00$\pm$0.02 &  1.05$\pm$0.01 & 1.06$\pm$0.03  &  1.04$\pm$0.04  &  0.76$\pm$0.02\\
 0.152 &   1.45 &  0.96 &   1.02$\pm$0.02 &  1.02$\pm$0.02 & 1.03$\pm$0.02  &  1.05$\pm$0.03  &  0.81$\pm$0.02\\
 0.124 &   1.37 &  0.94 &   1.02$\pm$0.02 &  1.06$\pm$0.02 & 1.09$\pm$0.03  &  1.12$\pm$0.05  &  0.81$\pm$0.02\\
 0.102 &   1.29 &  0.92 &   1.01$\pm$0.02 &  1.02$\pm$0.02 & 1.04$\pm$0.02  &  1.05$\pm$0.03  &  0.83$\pm$0.02\\
 0.084 &   1.20 &  0.89 &   0.99$\pm$0.02 &  1.03$\pm$0.02 & 1.06$\pm$0.03  &  1.04$\pm$0.04  &  0.86$\pm$0.02\\
 0.068 &   1.11 &  0.86 &   1.01$\pm$0.01 &  0.99$\pm$0.02 & 1.03$\pm$0.02  &  1.05$\pm$0.03  &  0.82$\pm$0.02\\
 0.055 &   1.00 &  0.81 &   0.97$\pm$0.02 &  1.00$\pm$0.02 & 1.05$\pm$0.04  &  1.11$\pm$0.05  &  0.87$\pm$0.03\\
 0.044 &   0.90 &  0.76 &   0.97$\pm$0.02 &  0.97$\pm$0.02 & 0.95$\pm$0.02  &  0.89$\pm$0.03  &  0.89$\pm$0.02\\
 0.035 &   0.78 &  0.69 &   0.98$\pm$0.02 &  0.95$\pm$0.02 &      -         &  0.94$\pm$0.05  &  0.89$\pm$0.02\\
 0.027 &   0.66 &  0.61 &   0.98$\pm$0.03 &  0.91$\pm$0.02 & 0.90$\pm$0.04  &  0.81$\pm$0.06  &  0.86$\pm$0.03\\
 0.020 &   0.54 &  0.51 &   0.99$\pm$0.03 &  0.89$\pm$0.03 & 0.85$\pm$0.05  &  0.87$\pm$0.08  &  1.03$\pm$0.04\\
 0.014 &   0.40 &  0.40 &   0.99$\pm$0.05 &  0.87$\pm$0.04 & 0.86$\pm$0.08  &  0.82$\pm$0.14  &  1.04$\pm$0.07\\
\hline 

\hline
\end{tabular}
\end{table}

\newpage

\end{document}